\documentclass[conference]{IEEEtran}
\IEEEoverridecommandlockouts

\usepackage{cite}
\usepackage{amsmath,amssymb,amsfonts}
\usepackage{algorithmic}
\usepackage{graphicx}
\usepackage{textcomp}
\usepackage{xcolor}
\usepackage[utf8]{inputenc}
\usepackage{subfigure}
\usepackage{placeins}
\usepackage[ruled,vlined]{algorithm2e}
\def\BibTeX{{\rm B\kern-.05em{\sc i\kern-.025em b}\kern-.08em
    T\kern-.1667em\lower.7ex\hbox{E}\kern-.125emX}}
\begin{document}

\title{Probability of Error Analysis for NOMA Systems in Rayleigh Fading Channels: Enabling IoT in Civil Engineering\\
}

\author{
	\IEEEauthorblockN{1\textsuperscript{st} Amr Abdelbari}
	\IEEEauthorblockA{\textit{Artificial Intelligence Engineering Dept.,} \\
		\textit{AI and Robotics Institute} \\
		\textit{Near East University} \\
		Nicosia, TRNC, Mersin 10, Turkey \\
		amr.abdelbari@neu.edu.tr}
}

\maketitle

\begin{abstract}
In the realm of digital communication, understanding and mitigating the probability of error is crucial, particularly in Rayleigh fading channels where signal impairments are common. This paper presents a unified approach to derive the probability of error formulations for two-users NOMA systems operating in Rayleigh fading channels. The methodologies and findings outlined in this study are essential for IoT applications in construction and civil engineering. Specifically, the derived error probability formulations can be employed to enhance the reliability and efficiency of IoT-based monitoring systems in these sectors. By optimizing communication protocols, the proposed approach ensures accurate data transmission, thereby facilitating real-time monitoring and decision-making processes in construction sites and civil infrastructure projects.
\end{abstract}

\begin{IEEEkeywords}
NOMA, Bit error rate (BER), probability of error, BPSK, QPSK
\end{IEEEkeywords}

\section{Introduction}\label{Introduction}
The advent of the Internet of Things (IoT) has revolutionized various sectors, including construction and civil engineering, by enabling smart and interconnected systems \cite{ADHIKARY20235404}. In these sectors, ensuring reliable communication is critical for real-time monitoring and efficient operations \cite{doi:10.1002/dac.4229}. Non-Orthogonal Multiple Access (NOMA) has emerged as a promising technology to enhance the capacity and efficiency of wireless communication systems, especially in challenging environments such as Rayleigh fading channels \cite{8246842}. NOMA distinguishes users by allocating different power levels to their signals. Users with better channel conditions (e.g., those closer to the base station) are assigned lower power levels, while users with poorer channel conditions (e.g., those farther from the base station) are assigned higher power levels. This approach ensures that all users can simultaneously access the same frequency resources without causing significant interference to each other \cite{9179825}. NOMA allows multiple IoT devices to share the same frequency resources simultaneously, increasing the number of devices that can be connected within a given bandwidth \cite{7842433}. NOMA supports low latency communication, which is crucial for real-time monitoring and control applications in IoT. This includes applications in construction, intelligent transportation systems, and real-time infrastructure monitoring \cite{ADHIKARY20235404}.

This paper delves into the intricacies of error probability in NOMA systems, specifically in the context of construction and civil engineering applications. By addressing the probability of error for two users NOMA system, this study provides a comprehensive framework that can significantly benefit IoT implementations in these fields \cite{9397776}.

\begin{figure}
	\centering
	\includegraphics[width=\linewidth]{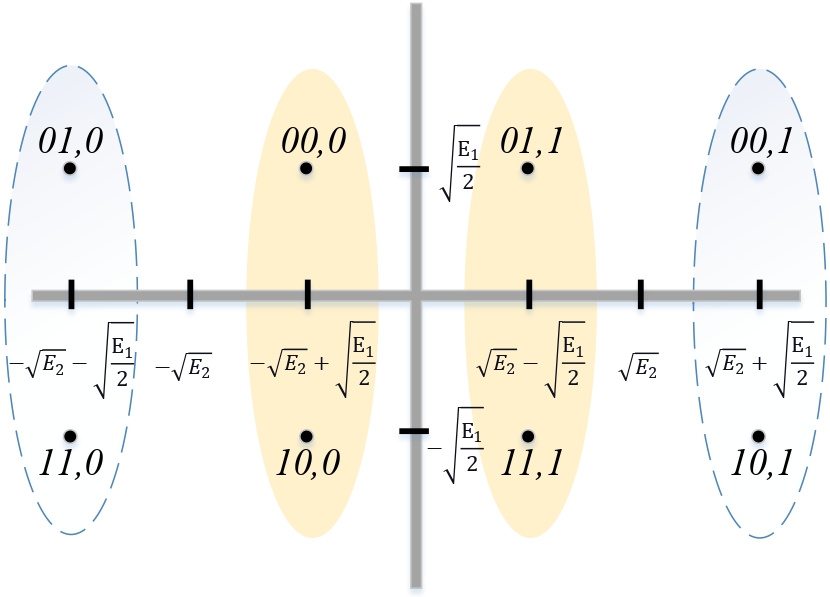}
	\caption{Constellation plot for two users (BPSK, QPSK) NOMA system.}
	\label{fig:bpskqpsknoma}
\end{figure}

\section{Downlink NOMA Model}

The downlink NOMA scenario involves one base station (BS) transmitting to two users over the same frequency and time slots using Superposition Coding (SC).
Received signal for the $k$-th user is expressed by
\begin{equation}\label{key}
y_k = h_k x_{sc} + w_k, \quad k = 1, 2
\end{equation}
where $h_k$ denotes the channel coefficient for the $k$-th user, representing Rayleigh fading, $x_{sc}$ is the superposition coded signal transmitted by the BS. $w_k$ is the Additive White Gaussian Noise (AWGN) at the receiver of the $k$-th user.
The superposition coded signal is
\begin{equation}\label{key}
	x_{sc} = \sqrt{\epsilon_1} x_1 + \sqrt{\epsilon_2} x_2,
\end{equation}
where \( \epsilon_1 \) and \( \epsilon_2 \) are power allocation coefficients for user 1 and user 2, respectively and \( x_1 \) and \( x_2 \) are modulated symbols for user 1 and user 2, respectively. The power allocation coefficients satisfy \( \epsilon_1 = \alpha P_s \) and \( \epsilon_2 = (1 - \alpha) P_s \), where \( \alpha \) is the power allocation factor and \( P_s \) is the total transmit power.
It is assumed that  User 1 is  the near user (NU) and thus has a better channel conditions ($h_1^2 > h_2^2$) while user 2 is far user (FU) and has worse channel conditions. Based on these assumptions, User 1 (NU) has higher priority and better channel quality while User 2 (FU) experiences more severe channel conditions.
On the detection process at User 2 (FU), User 2 treats the signal intended for User 1 as interference and directly detects its own signal using a Maximum-Likelihood (ML) detector.

On the other hand, user 1 (NU) imploies Successive Interference Cancellation (SIC) to detect his own signal. First, user 1 first detects the signal intended for User 2. After detecting \( x_2 \), User 1 subtracts the reconstructed signal from its received signal and then detects its own signal.
User 1 first detects \( x_2 \):
Then, subtracts \( \hat{x}_2 \) from \( y_1 \) and detects \( x_1 \).
The total error probability for User 1 is given as the sum of the error probabilities under two conditions: correct and erroneous detection of \( x_2 \).

\section{Performance Analysis}\label{proposed_technique}
For the FU Binary Phase Shift Keying (BPSK), the derivation begins by understanding the properties of  BPSK modulation within an AWGN channel. BPSK represents binary digits, 1 and 0, through analog levels $+\sqrt{E_b}$ and $-\sqrt{E_b}$ respectively. 
The FU uses the ML detector to minimize the probability of error. The error occurs if the noise causes the received signal to cross the decision boundary incorrectly.

For binary modulated signals (like BPSK for the FU), the decision boundary is the horizontal line at 0. For Quadrature Phase Shift Keying (QPSK) modulation (used in NU), the decision boundaries involve considering the in-phase and quadrature components separately.
The prior probability of a symbol is the likelihood that the symbol was transmitted before any observations. In most cases, for simplicity, it is assumed that all symbols are equally likely (uniform prior probability).
The total error probability of the ML detector is obtained by summing the error probabilities of all possible symbols, each weighted by their prior probabilities.
If symbols have equal prior probabilities, this weighting becomes a simple average ($1/8$) in our case \cite{2018.5278}.
For (10,0) and (00,0) at ($-\sqrt{\epsilon_2} + \sqrt{\dfrac{\epsilon_{1}}{2}}$), noise should be ($ n \ge\sqrt{\epsilon_2} - \sqrt{\dfrac{\epsilon_{1}}{2}}$) to be detected erroneously.
For (01,0) and (11,0) at ($-\sqrt{\epsilon_2} - \sqrt{\dfrac{\epsilon_{1}}{2}}$), noise should be ($ n \ge\sqrt{\epsilon_2} + \sqrt{\dfrac{\epsilon_{1}}{2}}$) to be detected erroneously.
For (01,1) and (11,1) at ($\sqrt{\epsilon_2} - \sqrt{\dfrac{\epsilon_{1}}{2}}$), noise should be ($ n \le - \sqrt{\epsilon_2} + \sqrt{\dfrac{\epsilon_{1}}{2}}$) to be detected erroneously.
For (00,1) and (10,1) at ($\sqrt{\epsilon_2} + \sqrt{\dfrac{\epsilon_{1}}{2}}$), noise should be ($ n \le -\sqrt{\epsilon_2} - \sqrt{\dfrac{\epsilon_{1}}{2}}$) to be detected erroneously \cite{2018.5278}.

For (00,0) and (10,0) transmission, the probability of error in FU is determined by the integral of the tail of the Gaussian distribution extending from 0 to $\infty$:
\begin{equation}\label{BPSK_BER_derivation}
	p_{2}(e|(00,0)) = \frac{1}{\sqrt{2\pi \sigma^2}} \int_{0}^{\infty} e^{-\frac{(y - (-\sqrt{\epsilon_2} + \sqrt{\dfrac{\epsilon_{1}}{2}})h_{2})^2}{2\sigma^2}} dy.
\end{equation}
Using a substitution where $z^{2} = \dfrac{(y - (-\sqrt{\epsilon_2} + \sqrt{\dfrac{\epsilon_{1}}{2}})h_{2})^2}{N_{0}}$ and $N_{0} = 2\sigma^{2}$. Thus, $z = \dfrac{y +\sqrt{\epsilon_2}h_{2} - \sqrt{\dfrac{\epsilon_{1}}{2}}h_{2}}{\sqrt{N_{0}}}$, $dz = \dfrac{1}{\sqrt{N_{0}}}dy$, and $dy = \sqrt{N_{0}}dz$. The integration limits will be as follows: $y = 0 \Rightarrow z = \dfrac{\sqrt{\epsilon_2}h_{2} - \sqrt{\dfrac{\epsilon_{1}}{2}}h_{2}}{\sqrt{N_{0}}}$ and $y = \infty \Rightarrow z = \infty$.
Equation (\ref{BPSK_BER_derivation}) simplifies to:
\begin{equation}
	\begin{split}
	p_{2}(e|(00,0)) = & \dfrac{1}{\sqrt{\pi}} \int_{\dfrac{\sqrt{\epsilon_2}h_{2} - \sqrt{\dfrac{\epsilon_{1}}{2}}h_{2}}{\sqrt{N_{0}}}}^{\infty} e^{-z^{2}} dz  \\
		= & \dfrac{1}{2} erfc(\dfrac{\sqrt{\epsilon_2}h_{2} - \sqrt{\dfrac{\epsilon_{1}}{2}}h_{2}}{\sqrt{N_{0}}}) \\
		= & \dfrac{2}{2} Q(\dfrac{\sqrt{\epsilon_2}h_{2} - \sqrt{\dfrac{\epsilon_{1}}{2}}h_{2}}{\sqrt{N_{0}}}\sqrt{2}) \\
				= & Q(\dfrac{\sqrt{\epsilon_2}h_{2} - \sqrt{\dfrac{\epsilon_{1}}{2}}h_{2}}{\sqrt{\dfrac{N_{0}}{2}}}) \\
	\end{split}
\end{equation}
where 
\begin{equation}
	erfc(x) =  \dfrac{2}{\sqrt{\pi}} \int_{x}^{\infty} e^{-x^{2}} dx.
\end{equation}

For (01,1) and (11,1) transmission, the probability of error in FU is determined by the integral of the tail of the Gaussian distribution extending from $-\infty$ to 0:
\begin{equation}\label{BPSK_BER_derivation_2}
	p_{2}(e|(01,1)) = \frac{1}{\sqrt{\pi N_{0}}} \int_{-\infty}^{0} e^{-\frac{(y - (\sqrt{\epsilon_2} - \sqrt{\dfrac{\epsilon_{1}}{2}})h_{2})^2}{2\sigma^2}} dy.
\end{equation}
Using a substitution where $z^{2} = \dfrac{(y - (+\sqrt{\epsilon_2} - \sqrt{\dfrac{\epsilon_{1}}{2}})h_{2})^2}{N_{0}}$ and $N_{0} = 2\sigma^{2}$. Thus, $z = \dfrac{y -\sqrt{\epsilon_2}h_{2} +\sqrt{\dfrac{\epsilon_{1}}{2}}h_{2}}{\sqrt{N_{0}}}$, $dz = \dfrac{1}{\sqrt{N_{0}}}dy$, and $dy = \sqrt{N_{0}}dz$. The integration limits will be as follows: $y = 0 \Rightarrow z = \dfrac{-\sqrt{\epsilon_2}h_{2} + \sqrt{\dfrac{\epsilon_{1}}{2}}h_{2}}{\sqrt{N_{0}}}$ and $y = -\infty \Rightarrow z = -\infty$.
Equation (\ref{BPSK_BER_derivation_2}) simplifies to:
\begin{equation}
	\begin{split}
		p_{2}(e|(01,1)) = & \dfrac{1}{\sqrt{\pi}} \int_{\dfrac{\sqrt{\epsilon_2}h_{2} - \sqrt{\dfrac{\epsilon_{1}}{2}}h_{2}}{\sqrt{N_{0}}}}^{\infty} e^{-z^{2}} dz  \\
		= & \dfrac{1}{2} erfc(\dfrac{\sqrt{\epsilon_2}h_{2} - \sqrt{\dfrac{\epsilon_{1}}{2}}h_{2}}{\sqrt{N_{0}}}) \\
		= & Q(\dfrac{\sqrt{\epsilon_2}h_{2} - \sqrt{\dfrac{\epsilon_{1}}{2}}h_{2}}{\sqrt{\dfrac{N_{0}}{2}}}). \\
	\end{split}
\end{equation}

The same way for (01,0) and (11,0),
\begin{equation}
	\begin{split}
		p_{2}(e|(01,0)) 
		= & \dfrac{1}{2} erfc(\dfrac{\sqrt{\epsilon_2}h_{2} + \sqrt{\dfrac{\epsilon_{1}}{2}}h_{2}}{\sqrt{N_{0}}}) \\
		= & Q(\dfrac{\sqrt{\epsilon_2}h_{2} + \sqrt{\dfrac{\epsilon_{1}}{2}}h_{2}}{\sqrt{\dfrac{N_{0}}{2}}}). \\
	\end{split}
\end{equation}
which is the same for (00,1) and (10,1),
\begin{equation}
	\begin{split}
		p_{2}(e|(00,1))
		= & \dfrac{1}{2} erfc(\dfrac{\sqrt{\epsilon_2}h_{2} + \sqrt{\dfrac{\epsilon_{1}}{2}}h_{2}}{\sqrt{N_{0}}}) \\
		= & Q(\dfrac{\sqrt{\epsilon_2}h_{2} + \sqrt{\dfrac{\epsilon_{1}}{2}}h_{2}}{\sqrt{\dfrac{N_{0}}{2}}}). \\
	\end{split}
\end{equation}

The total BER of the FU is:
\begin{equation}
	P_{b, 2} = \sum_{i=0}^{MN-1}p(s_{i})p(e|s_{i})
\end{equation}
where $s_{i}$ is the $i$th bit in the superposition signal of two users with modulation M and N. Here M = 4 for NU and N = 2 for FU. Assuming equal probabilities for the 8 points (1/8), the total BER is 
\begin{equation}\label{BER_BPSK}
	P_{b, 2} = \dfrac{1}{2} [Q(\dfrac{\sqrt{\epsilon_2}h_{2} + \sqrt{\dfrac{\epsilon_{1}}{2}}h_{2}}{\sqrt{\dfrac{N_{0}}{2}}}) + Q(\dfrac{\sqrt{\epsilon_2}h_{2} - \sqrt{\dfrac{\epsilon_{1}}{2}}h_{2}}{\sqrt{\dfrac{N_{0}}{2}}})].
\end{equation}
Let $\gamma_{A} = \frac{(\sqrt{\epsilon_2}- \sqrt{\dfrac{\epsilon_{1}}{2}})^{2}|h_{2}|^{2}}{\dfrac{N_{0}}{2}}$ and $\gamma_{B} = \frac{(\sqrt{\epsilon_2} + \sqrt{\dfrac{\epsilon_{1}}{2}})^{2}|h_{2}|^{2}}{\dfrac{N_{0}}{2}}$, then:
\begin{equation}\label{}
		\begin{split}
	P_{b, 2} = & \dfrac{1}{2} [Q(\sqrt{\gamma_{A}}) + Q(\sqrt{\gamma_{B}})] \\
		= & \dfrac{1}{4} [erfc(\sqrt{\dfrac{\gamma_{A}}{2}}) + erfc(\sqrt{\dfrac{\gamma_{B}}{2}})].
\end{split}
\end{equation}

In Rayleigh fading channel, the average BER of user 2 FU is:

\begin{equation}\label{BER_Rayleigh}
	\begin{split}
		P_{b, 2} = & \dfrac{1}{2} [\int_{0}^{\infty}Q(\sqrt{\gamma_{A}})f_{\gamma_{A}}(\gamma_{A})d\gamma_{A} + \int_{0}^{\infty}Q(\sqrt{\gamma_{B}})f_{\gamma_{B}}(\gamma_{B})d\gamma_{B}] \\
		= &\dfrac{1}{4} [\int_{0}^{\infty}erfc(\sqrt{\dfrac{\gamma_{A}}{2}})f_{\gamma_{A}}(\gamma_{A})d\gamma_{A} +\\ & \int_{0}^{\infty}erfc(\sqrt{\dfrac{\gamma_{B}}{2}})f_{\gamma_{B}}(\gamma_{B})d\gamma_{B}].
	\end{split}
\end{equation}
where 
The probability density of the magnitude $|h|$ is,
\begin{equation}\label{gamma}
	\begin{split}
		p(\gamma) = & \frac{1}{E_{b}/N_{0}}e^{\frac{-\gamma}{E_{b}/N_{0}}}, ~ \gamma \ge 0 \\
		= & \frac{1}{\bar{\gamma}}e^{\frac{-\gamma}{\bar{\gamma}}} \\
	\end{split}
\end{equation}
where $\bar{\gamma} = E_{b}/N_{0} = \bar{\gamma_{A}}/2$.
By substituting (\ref{gamma}) into the first part of (\ref{BER_Rayleigh}) is
\begin{equation}\label{BER_Rayleigh_1st_part}
		P_{b, 2}(1) 
		= \dfrac{1}{4\bar{\gamma}} \int_{0}^{\infty}erfc(\sqrt{\dfrac{\gamma_{A}}{2}})e^{\frac{-\gamma_{A}}{\bar{\gamma}}} d\gamma_{A}.
\end{equation}
 From (22) in \cite{abdelbari2024derivationprobabilityerrorbpsk, 201463}, let $\alpha = \bar{\gamma} = \dfrac{\bar{\gamma_{A}}}{2}$ and $x = \dfrac{\gamma_{A}}{2}$, then
\begin{equation}\label{helping_erfc_intergral_2}
	\begin{split}
		\int erfc\big(\sqrt{x}\big)e^{-\frac{x}{\alpha}}dx = & -\alpha erfc(\sqrt{x})e^{-\frac{x}{\alpha}} - \\
		& \alpha(\frac{\alpha}{\alpha+1})^{\frac{1}{2}} erf(\sqrt{\frac{(\alpha + 1)}{\alpha}x})\\
		= & -\dfrac{\bar{\gamma_{A}}}{2} erfc(\sqrt{\dfrac{\gamma_{A}}{2}})e^{-\frac{\dfrac{\gamma_{A}}{2}}{\dfrac{\bar{\gamma_{A}}}{2}}} - \\
		& \dfrac{\bar{\gamma_{A}}}{2}\sqrt{\frac{\dfrac{\bar{\gamma_{A}}}{2}}{\dfrac{\bar{\gamma_{A}}}{2}+1}} erf(\sqrt{\frac{(\dfrac{\bar{\gamma_{A}}}{2}+ 1)}{\dfrac{\bar{\gamma_{A}}}{2}}\dfrac{\gamma_{A}}{2}})\big|^{0}_{\infty}.\\
	\end{split}
\end{equation}
Since  $erfc(0) = 1$, $erfc(\infty) = 0$, $erf(0) = 0$ and $erf(\infty) = 1$, (\ref{BER_Rayleigh_1st_part}) will be
\begin{equation}\label{}
	\begin{split}
	P_{b, 2}(1) 
	= &\dfrac{1}{4\dfrac{\bar{\gamma_{A}}}{2}} [ \dfrac{\bar{\gamma_{A}}}{2} -  \dfrac{\bar{\gamma_{A}}}{2}\sqrt{\frac{\dfrac{\bar{\gamma_{A}}}{2}}{\dfrac{\bar{\gamma_{A}}}{2}+1}}] \\
	= &\dfrac{1}{4} [ 1- \sqrt{\frac{\bar{\gamma_{A}}}{\bar{\gamma_{A}}+2}}].\\
\end{split}
\end{equation}
The same for the second part of (\ref{BER_Rayleigh}) 
\begin{equation}\label{}
		P_{b, 2}(2) 
		= \dfrac{1}{4} [ 1- \sqrt{\frac{\bar{\gamma_{B}}}{\bar{\gamma_{B}}+2}}].
\end{equation}
Equation (\ref{BER_Rayleigh}) will be
\begin{equation}\label{}
	\begin{split}
		P_{b, 2}
		= &\dfrac{1}{4} [ (1- \sqrt{\frac{\bar{\gamma_{A}}}{\bar{\gamma_{A}}+2}})\\
		& + ( 1- \sqrt{\frac{\bar{\gamma_{B}}}{\bar{\gamma_{B}}+2}})].
	\end{split}
\end{equation}

On the other hand for user 1 (NU with QPSK and low power) the BER is 
\begin{equation}\label{user_1_BER}
	P_{1,e} = P_{1,e}^{\text{correct}} + P_{1,e}^{\text{error}}.
\end{equation}
From now on, ($b_{2,2}b_{2,1}, b_{1,1}$) is the symbol where the 1st number is the $k$th user and 2nd number is the bit order.
The FU bit detection affects the detection of the NU 1st bit. For 2nd bit, it does not matter if FU detected correctly  or erroneously. For the detection of NU 1st bit, there are two classes. Class I when the FU bit is correctly decoded and class II when the FU bit is erroneously decoded. 
In class I, the FU bit to be decoded correctly, the noise should not exceed the limit boundary which is horizontally at 0.
For (10,0) and (00,0) at ($-\sqrt{\epsilon_2} + \sqrt{\dfrac{\epsilon_{1}}{2}}$),
noise should be less than ($+\sqrt{\epsilon_2} - \sqrt{\dfrac{\epsilon_{1}}{2}}$) for the 1st bit of FU to be correctly detected and noise should be ($ n_{I} \le - \sqrt{\dfrac{\epsilon_{1}}{2}}$) for the 1st bit of NU to be detected erroneously. This puts the boundaries at ($ - \sqrt{\dfrac{\epsilon_{1}}{2}}$).
For (01,0) and (11,0) at ($-\sqrt{\epsilon_2} - \sqrt{\dfrac{\epsilon_{1}}{2}}$), to detect the 1st bit of FU correctly, ($n_{I} < +\sqrt{\epsilon_2} + \sqrt{\dfrac{\epsilon_{1}}{2}}$).
Therefore, for the 1st bit of NU, noise should be ($ n_{I} \leq \sqrt{\epsilon_2} + \sqrt{\dfrac{\epsilon_{1}}{2}}$) and  ($ n_{I} \ge \sqrt{\dfrac{\epsilon_{1}}{2}}$)  to be detected erroneously.
In the same way, for (01,1) and (11,1) at ($\sqrt{\epsilon_2} - \sqrt{\dfrac{\epsilon_{1}}{2}}$), noise should be ($ n_{I} \ge \sqrt{\dfrac{\epsilon_{1}}{2}}$) to be detected erroneously.
Finally, for (00,1) and (10,1) at ($\sqrt{\epsilon_2} + \sqrt{\dfrac{\epsilon_{1}}{2}}$), noise should be ($ n_{I} \ge -\sqrt{\epsilon_2} - \sqrt{\dfrac{\epsilon_{1}}{2}}$) and ($ n_{I} \le - \sqrt{\dfrac{\epsilon_{1}}{2}}$)  to be detected erroneously.
Accordingly, we have two probabilities Pr($\sqrt{\dfrac{\epsilon_{1}}{2}} \leq |n_{I}| \leq \sqrt{\epsilon_2} - \sqrt{\dfrac{\epsilon_{1}}{2}}$) and Pr($\sqrt{\dfrac{\epsilon_{1}}{2}} \leq |n_{I}|$).
For the 2nd bit of NU, we have Pr($\sqrt{\dfrac{\epsilon_{1}}{2}} \leq |n_{Q}|$) multiplied by the probability of detecting the FU correctly and erroneously \cite{7967766}.
\begin{equation}\label{}
	\begin{split}
			P_{b, 1| \text{correctFU}} = &\dfrac{1}{4} [Pr(\sqrt{\dfrac{\epsilon_{1}}{2}} \leq |n_{I}|) + \\ 
			& Pr(\sqrt{\dfrac{\epsilon_{1}}{2}} \leq |n_{I}| \leq \sqrt{\epsilon_2} - \sqrt{\dfrac{\epsilon_{1}}{2}}) + \\
			& Pr(\sqrt{\dfrac{\epsilon_{1}}{2}} \leq |n_{Q}|) * \\
			& (Pr(|n_{I}| \leq \sqrt{\epsilon_2} - \sqrt{\dfrac{\epsilon_{1}}{2}}) + \\
			& Pr(n_{I}| \leq \sqrt{\epsilon_2} + \sqrt{\dfrac{\epsilon_{1}}{2}}))].
	\end{split}
\end{equation}

The \(\Phi\) function and the \(Q\) function are both related to the standard normal distribution.
The \(\Phi\) function, also known as the cumulative distribution function (CDF) of the standard normal distribution, is defined as:
\begin{equation}\label{}
	\Phi(z) = P(Z \le z) = \frac{1}{\sqrt{2\pi}} \int_{-\infty}^{z} e^{-\frac{t^2}{2}} \, dt
\end{equation}
where \(Z\) is a standard normal random variable with mean 0 and variance 1. The \(\Phi\) function gives the probability that a standard normal random variable is less than or equal to \(z\).
The \(Q\) function is the complementary cumulative distribution function (CCDF) of the standard normal distribution. It is defined as:
\begin{equation}\label{}
	Q(z) = P(Z > z) = 1 - \Phi(z) = \frac{1}{\sqrt{2\pi}} \int_{z}^{\infty} e^{-\frac{t^2}{2}} \, dt
\end{equation}
This property is useful because it allows us to convert between the lower tail probabilities and the upper tail probabilities of the standard normal distribution.
Solving using \(\Phi\) function and the Q-function as in (\ref{BPSK_BER_derivation}), the BER of NU when the FU is correctly decoded will be
\begin{equation}\label{}
	\begin{split}
		P_{b, 1| \text{correctFU}} = & \dfrac{1}{4} \big[Q(\sqrt{\dfrac{\epsilon_{1}}{N_{0}}}h_{1})+ \\
		&Q(\sqrt{\dfrac{\epsilon_{1}}{N_{0}}}h_{1})-  Q(\dfrac{\sqrt{\epsilon_2}h_{1} + \sqrt{\dfrac{\epsilon_{1}}{2}}h_{1}}{\sqrt{\dfrac{N_{0}}{2}}}) +  \\
		& Q(\sqrt{\dfrac{\epsilon_{1}}{N_{0}}}h_{1})* \big[\big(1- Q(\dfrac{\sqrt{\epsilon_2}h_{1} + \sqrt{\dfrac{\epsilon_{1}}{2}}h_{1}}{\sqrt{\dfrac{N_{0}}{2}}}) \big) + \\ 
		& \big(1 - Q(\dfrac{\sqrt{\epsilon_2}h_{1	} - \sqrt{\dfrac{\epsilon_{1}}{2}}h_{1}}{\sqrt{\dfrac{N_{0}}{2}}})\big)\big]\big].
	\end{split}
\end{equation}
Let $\gamma_{C} = \dfrac{\epsilon_{1}}{N_{0}}|h_{1}|^{2}$,  $\gamma_{D} = \frac{(\sqrt{\epsilon_2}+ \sqrt{\dfrac{\epsilon_{1}}{2}})^{2}|h_{1}|^{2}}{\dfrac{N_{0}}{2}}$ and $\gamma_{E} = \frac{(\sqrt{\epsilon_2} - \sqrt{\dfrac{\epsilon_{1}}{2}})^{2}|h_{1}|^{2}}{\dfrac{N_{0}}{2}}$, then:
\begin{equation}\label{}
	\begin{split}
		P_{b, 1|\text{correctFU}} = & \dfrac{1}{4} [Q(\sqrt{\gamma_{C}}) + Q(\sqrt{\gamma_{C}}) - Q(\sqrt{\gamma_{D}}) + \\
		& Q(\sqrt{\gamma_{C}})* [2 - Q(\sqrt{\gamma_{D}}) - Q(\sqrt{\gamma_{E}})]] \\
		= & \dfrac{1}{4} [Q(\sqrt{\gamma_{C}})* [4 - Q(\sqrt{\gamma_{D}}) - \\ 
		& Q(\sqrt{\gamma_{E}})]- Q(\sqrt{\gamma_{D}}].
	\end{split}
\end{equation}

For class II  where an error happened in the detection of FU bit, the error in the 1st bit of NU is as following for each pair of symbols.
For (10,0) and (00,0) at ($-\sqrt{\epsilon_2} + \sqrt{\dfrac{\epsilon_{1}}{2}}$),
noise should be higher than ($+\sqrt{\epsilon_2} - \sqrt{\dfrac{\epsilon_{1}}{2}}$) for the 1st bit of FU to be erroneously detected and noise should be ($ n_{I} \le 2\sqrt{\epsilon_2} - \sqrt{\dfrac{\epsilon_{1}}{2}}$) for the 1st bit of NU to be detected erroneously.
For (01,0) and (11,0) at ($-\sqrt{\epsilon_2} - \sqrt{\dfrac{\epsilon_{1}}{2}}$), to detect the 1st bit of FU erroneously, ($n_{I} \ge +\sqrt{\epsilon_2} + \sqrt{\dfrac{\epsilon_{1}}{2}}$).
Therefore, for the 1st bit of NU, noise should be ($ n_{I} \ge \sqrt{\epsilon_2} + \sqrt{\dfrac{\epsilon_{1}}{2}}$) and  ($ n_{I} \ge 2\sqrt{\epsilon_2} + \sqrt{\dfrac{\epsilon_{1}}{2}}$)  to be detected erroneously.
In the same way, for (01,1) and (11,1) at ($\sqrt{\epsilon_2} - \sqrt{\dfrac{\epsilon_{1}}{2}}$), noise should be ($ n_{I} \le -\sqrt{\epsilon_2} + \sqrt{\dfrac{\epsilon_{1}}{2}}$) and  ($ n_{I} \ge -2\sqrt{\epsilon_2} + \sqrt{\dfrac{\epsilon_{1}}{2}}$)  to be detected erroneously.
Finally, for (00,1) and (10,1) at ($\sqrt{\epsilon_2} + \sqrt{\dfrac{\epsilon_{1}}{2}}$), noise should be ($ n_{I} \le -\sqrt{\epsilon_2} - \sqrt{\dfrac{\epsilon_{1}}{2}}$) and ($ n_{I} \le -2\sqrt{\epsilon_2} - \sqrt{\dfrac{\epsilon_{1}}{2}}$)  to be detected erroneously.
Accordingly, we have two probabilities Pr($\sqrt{\epsilon_2} - \sqrt{\dfrac{\epsilon_{1}}{2}}< |n_{I}| < 2\sqrt{\epsilon_2} - \sqrt{\dfrac{\epsilon_{1}}{2}}$) and Pr($2\sqrt{\epsilon_2} + \sqrt{\dfrac{\epsilon_{1}}{2}} < |n_{I}|$).
As in class I, for the 2nd bit of NU, we have Pr($\sqrt{\dfrac{\epsilon_{1}}{2}} < |n_{Q}|$) multiplied by the probability of detecting the FU correctly and erroneously.
\begin{equation}\label{}
	\begin{split}
		P_{b, 1| \text{errorFU}} = &\dfrac{1}{4} [Pr(2\sqrt{\epsilon_2} + \sqrt{\dfrac{\epsilon_{1}}{2}} \le |n_{I}|) + \\ 
		& Pr(\sqrt{\epsilon_2} - \sqrt{\dfrac{\epsilon_{1}}{2}}\le |n_{I}| < 2\sqrt{\epsilon_2} - \sqrt{\dfrac{\epsilon_{1}}{2}}) + \\
		& Pr(\sqrt{\dfrac{\epsilon_{1}}{2}} \le |n_{Q}|) * \\
		& (Pr( |n_{I}| \ge \sqrt{\epsilon_2} - \sqrt{\dfrac{\epsilon_{1}}{2}}) + \\
		& Pr(|n_{I}| \ge \sqrt{\epsilon_2} + \sqrt{\dfrac{\epsilon_{1}}{2}}))].
	\end{split}
\end{equation}
Solving using \(\Phi\) function and the Q-function as in (\ref{BPSK_BER_derivation}), the BER of NU when the FU is correctly decoded will be
\begin{equation}\label{}
	\begin{split}
		P_{b, 1| \text{errorFU}} = & \dfrac{1}{4} \big[Q(\dfrac{2\sqrt{\epsilon_2}h_{1} + \sqrt{\dfrac{\epsilon_{1}}{2}}h_{1}}{\sqrt{\dfrac{N_{0}}{2}}})+ \\
		& Q(\dfrac{\sqrt{\epsilon_2}h_{1} - \sqrt{\dfrac{\epsilon_{1}}{2}}h_{1}}{\sqrt{\dfrac{N_{0}}{2}}}) -  Q(\dfrac{2\sqrt{\epsilon_2}h_{1} - \sqrt{\dfrac{\epsilon_{1}}{2}}h_{1}}{\sqrt{\dfrac{N_{0}}{2}}}) +  \\
		& Q(\sqrt{\dfrac{\epsilon_{1}}{N_{0}}}h_{1})* \big[Q(\dfrac{\sqrt{\epsilon_2}h_{1} + \sqrt{\dfrac{\epsilon_{1}}{2}}h_{1}}{\sqrt{\dfrac{N_{0}}{2}}})+ \\ 
		& Q(\dfrac{\sqrt{\epsilon_2}h_{1} - \sqrt{\dfrac{\epsilon_{1}}{2}}h_{1}}{\sqrt{\dfrac{N_{0}}{2}}})\big]\big].
	\end{split}
\end{equation}
Let $\gamma_{F} = \frac{(2\sqrt{\epsilon_2}+ \sqrt{\dfrac{\epsilon_{1}}{2}})^{2}|h_{1}|^{2}}{\dfrac{N_{0}}{2}}$ and $\gamma_{G} = \frac{(2\sqrt{\epsilon_2} - \sqrt{\dfrac{\epsilon_{1}}{2}})^{2}|h_{1}|^{2}}{\dfrac{N_{0}}{2}}$, then:
\begin{equation}\label{}
	\begin{split}
		P_{b, 1|\text{errorFU}} = & \dfrac{1}{4} [Q(\sqrt{\gamma_{F}}) + Q(\sqrt{\gamma_{E}}) - Q(\sqrt{\gamma_{G}}) + \\
		& Q(\sqrt{\gamma_{C}})* [Q(\sqrt{\gamma_{D}}) + Q(\sqrt{\gamma_{E}})]]. \\
	\end{split}
\end{equation}

\begin{equation}\label{}
	\begin{split}
		P_{b, 1} = & P_{1,e|\text{correctFU}} + P_{1,e|\text{errorFU}} \\
		= & \dfrac{1}{4} [Q(\sqrt{\gamma_{C}})* [4 - Q(\sqrt{\gamma_{D}}) - Q(\sqrt{\gamma_{E}})]- Q(\sqrt{\gamma_{D}} +\\ 
		&  Q(\sqrt{\gamma_{F}}) + Q(\sqrt{\gamma_{E}}) - Q(\sqrt{\gamma_{G}}) + \\
		& Q(\sqrt{\gamma_{C}})* [Q(\sqrt{\gamma_{D}}) + Q(\sqrt{\gamma_{E}})]] \\
		= & \dfrac{1}{4} [Q(\sqrt{\gamma_{F}}) + Q(\sqrt{\gamma_{E}}) - Q(\sqrt{\gamma_{D}}) - Q(\sqrt{\gamma_{G}})] + \\
		& Q(\sqrt{\gamma_{C}}).
	\end{split}
\end{equation}
The total average BER of user 1 NU is averaged in Rayleigh fading channel as follows:
\begin{equation}\label{NU_BER_Rayleigh}
	\begin{split}
		P_{b, 1} = &  \dfrac{1}{4} [\int_{0}^{\infty}Q(\sqrt{\gamma_{F}})f_{\gamma_{F}}(\gamma_{F})d\gamma_{F}  +  \int_{0}^{\infty}Q(\sqrt{\gamma_{E}})f_{\gamma_{E}}(\gamma_{E})d\gamma_{E}  - \\ & \int_{0}^{\infty}Q(\sqrt{\gamma_{D}})f_{\gamma_{D}}(\gamma_{D})d\gamma_{D}  - \int_{0}^{\infty}Q(\sqrt{\gamma_{G}})f_{\gamma_{G}}(\gamma_{G})d\gamma_{G} ] + \\
		& \int_{0}^{\infty}Q(\sqrt{\gamma_{C}})f_{\gamma_{C}}(\gamma_{C})d\gamma_{C}. 
	\end{split}
\end{equation}
In a similar way to the solution of (\ref{BER_Rayleigh}), the average BER over Rayleigh fading for NU in (\ref{NU_BER_Rayleigh})  becomes
\begin{equation}\label{}
	\begin{split}
		P_{b, 1}
		= &\dfrac{1}{2}(1- \sqrt{\frac{\bar{\gamma_{C}}}{\bar{\gamma_{C}}+2}}) + \\
		& \dfrac{1}{8}[(1- \sqrt{\frac{\bar{\gamma_{E}}}{\bar{\gamma_{E}}+2}}) + ( 1- \sqrt{\frac{\bar{\gamma_{F}}}{\bar{\gamma_{F}}+2}}) - \\
		& ( 1- \sqrt{\frac{\bar{\gamma_{D}}}{\bar{\gamma_{D}}+2}}) - ( 1- \sqrt{\frac{\bar{\gamma_{G}}}{\bar{\gamma_{G}}+2}}) ] \\
		= &\dfrac{1}{2}(1- \sqrt{\frac{\bar{\gamma_{C}}}{\bar{\gamma_{C}}+2}}) + \dfrac{1}{8}[\sqrt{\frac{\bar{\gamma_{D}}}{\bar{\gamma_{D}}+2}} + \\
		&  \sqrt{\frac{\bar{\gamma_{G}}}{\bar{\gamma_{G}}+2}} - \sqrt{\frac{\bar{\gamma_{E}}}{\bar{\gamma_{E}}+2}} -  \sqrt{\frac{\bar{\gamma_{F}}}{\bar{\gamma_{F}}+2}}].
	\end{split}
\end{equation}

\section{Conclusion}
The research presented in this paper underscores the significance of accurately determining the probability of error for NOMA systems in Rayleigh fading channels. By developing a unified approach for BPSK, 16-QAM, and 64-QAM, this study provides valuable insights that are directly applicable to IoT implementations in construction and civil engineering. The enhanced communication reliability afforded by these formulations can lead to more robust and efficient IoT-based monitoring and control systems. As a result, construction sites and civil engineering projects can benefit from improved real-time data acquisition and analysis, leading to better project management and operational efficiency. Future research should focus on extending these findings to other modulation schemes and exploring their practical applications in diverse IoT environments within the construction and civil engineering domains.
\appendices


\bibliographystyle{IEEEtran}
\bibliography{IEEEabrv,main}

\end{document}